\documentclass[preprint,review,11pt]{elsarticle}
\usepackage{graphicx}
\usepackage{amssymb}

\newcommand{\pms}{$\pm$}
\newcommand{\cms}{cm$^{-2}$}
\newcommand{\D}{D$_{25}\,$}

\newcommand{\lum}{erg\, s$^{-1}$}

\newcommand{\Msun}{\ensuremath{M_{\odot}}}

\newcommand{\mnras}{MNRAS}

\newcommand{\tim}{$\times$}
\newcommand{\s}{$\sim$}
\newcommand{\p}{$\%$}
\newcommand{\lx}{L$_X$}
\newcommand{\ts}{10$^{37}$}
\newcommand{\te}{10$^{38}$}
\newcommand{\tn}{10$^{39}$}

\journal{New Astronomy}
\begin{document}
\begin{frontmatter}
\title{Spectral properties of XRBs in dusty early-type galaxies}
\author[]{N. D. Vagshette}
\author[]{M. B. Pandge}
\author{M.K.~Patil\corref{cor1}}
\ead{patil@iucaa.ernet.in}
\address{School of Physical Sciences, S.R.T.M. University, Nanded-431 606 (MS), India Tel.+91-2462-229242; +91-9405938449/Fax:+91-2462-229245}
\cortext[cor1]{Corresponding author}

\begin{abstract}
We present spectral properties of a total of 996 discrete X-ray sources resolved in a sample of 23 dusty early-type galaxies selected from different environments. The combined X-ray luminosity function of all the 996 sources within the optical \D\,of the sample galaxies is well described by a broken power law with a break at 2.71$\times$\te \lum\, and is close to the Eddington limit for a 1.4\Msun\,neutron star. Out of the 996, about 63\p\, of the sources have their X-ray luminosities in the range between few\tim\ts to 2.0 \tim \tn \lum and are like normal LMXBs; about 15-20\p\, with luminosities $<$ few \tim 10$^{37}$ \lum\, are either super-soft or very-soft sources; while the remainder represents ULXs, HMXBs or unrelated heavily absorbed harder sources. More XRBs have been detected in the galaxies from isolated regions while those from rich groups and clusters host very few sources. The X-ray color-color plot for these sources has enabled us to classify them as SNRs, LMXBs, HMXBs and heavily absorbed AGNs. The composite X-ray spectra of the resolved sources within \D\, region of each of the galaxies are best represented by a power law with the average photon spectral index close to 1.65. The contribution of the resolved sources to the total X-ray luminosity of their host is found to vary greatly, in the sense that, in galaxies like NGC 3379 the XRB contribution is about 81\p\, while for NGC 5846 it is only 2\p. A correlation has been evidenced between the cumulative X-ray luminosity of the resolved sources against the star formation rate and the Ks band luminosity of the target galaxies indicating their primordial origin.
\end{abstract}
\end{frontmatter}
\section{Introduction}
The X-ray emission from early-type galaxies (ETGs; ellipticals and lenticulars)  has been subject of investigations and theoretical modeling ever since their detection for the first time with the \textit{Einstein} telescope (\cite{1985ApJ...293..102F}). It is now well established that X-rays from these galaxies partly originates from hot interstellar medium (ISM) and partly from a population of X-ray binaries (XRBs; see e.g. reviews by \cite{1989ApJ...347..127F}, \cite{2006ARA&A..44..323F}), with the latter component being more important for the case of X-ray faint early-type galaxies (\cite{2000PASJ...52..685M}). An XRB contains either a neutron star (NS) or a black hole (BH) accreting material from a companion star. Regardless the nature of the compact object, depending on the mass of the donor star, XRBs come in two basic flavors: the high-mass X-ray binaries (HMXBs) and the low-mass X-ray binaries (LMXBs). HMXBs are predominantly powered by the stellar winds from a young massive O or B star and are mostly found in late-type galaxies. LMXBs are the long lived systems in which a Roche Lobe overflowing low mass (\s\, 1\Msun) star is supplying material to the compact accreting object. As early-type galaxies are dominated by the old stellar populations (with age $\ge$ 1\,Gyr), the majority of the X-ray sources in these systems are believed to be LMXBs rather than short lived (\s\,10$^7$ yr) high-mass objects. Thus, the population of HMXBs scales with the star-formation rate (\cite{2003MNRAS.339..793G}), while that of the LMXBs scales with the stellar content of the host galaxies (\cite{2004MNRAS.349..146G}).

\begin{table}
\caption{Global parameters of the program galaxies}
\begin{center}
\scriptsize{
\begin{tabular}{|c c c c c c c c c |}
\hline
Sr. &  Obj.       &	Morph.     &      z     &   Mag    & Obs-ID & Exp. & GTI  & Environment \\
No. &  	          &	           &            &  (M$_B$) &	    & (ks) & (ks) &         \\
(1) &   (2)	  &    (3)         &     (4)	&  (5)	   &  (6)   & (7) & (8) & (9)    \\
\hline	
 1&  NGC 1395  &	E2	  &    0.00573 &  10.97   &  	799   &  	28	&  	22      & BGG	\\
 2&  NGC 1399  &	E1	  &    0.00475 &  10.00   &  	240   &  	44	&  	43      & BCG	\\
 3&  NGC 1404  &	E1	  &    0.00649 &  10.95   &  	2942  &  	30	&  	29      & CG    \\
 4&  NGC 1407  &	E0	  &    0.00593 &  10.7	   &  	791   &  	49	&  	44.4    & CG	\\
 5&  NGC 2768  &        S0        &    0.00458 &  10.84   &  	9528  &  	65	&  	64      & FG	\\ 
 6&  NGC 3377  &        E5-6      &    0.00222 &  11.24   &  	2934  &  	40	&  	39.5    & GG    \\
 7&  NGC 3379  &	E1	  &    0.00304 &  10.24   &  	1587  &  	32	&  	30.9    & FG	\\
 8&  NGC 3585  &	E7/S0	  &    0.00478 &  10.88   &  	9506  &  	60	&  	59      & GG   \\
 9&  NGC 3607  &        SA(s)0	  &    0.0032  &  10.82   &  	2073  &  	39	&  	38.4    & BGG	\\
10&  NGC 3801  &        S0/a	  &    0.0111  &  12.96   &  	6843  &  	60	&  	58.2    & CG	\\
11&  NGC 3923  &	E7/S0	  &    0.00580 &  10.88   &  	1563  &  	22	&  	15.60   & BGG 	\\
12&  NGC 4125  &        E6	  &    0.00452 &  10.65   &  	2071  &  	65	&  	62      & BGG	\\
13&  NGC 4278  &        E1-2      &    0.00217 &  11.20   &  	7081  &  	111	&  	110     & GG    \\
14&  NGC 4365  &        E3        &    0.00415 &  10.52   &  	2015  &  	41	&  	40.4    & CG    \\
15&  NGC 4374  &	E1	  &    0.00354 &  10.09   &  	803   &  	29	&  	27.8    & CG	\\
16&  NGC 4473  &	E5	  &    0.00749 &  11.16   &  	4688  &  	30	&  	29.4    & CG	\\
17&  NGC 4494  &	E1-2	  &    0.00448 &  10.71   &  	2079  &  	25	&  	22      & GG	\\
18&  NGC 4552  &	E	  &    0.00113 &  10.73   &  	2072  &  	55	&  	53.7    & CG 	\\
19&  NGC 4649  &	E2	  &    0.00373 &  9.81	   &  	8182  &  	53	&  	48.8    & CG 	\\
20&  NGC 4697  &	E6	  &    0.00414 &  10.14   &  	784   &  	40	&  	38      & BGG	\\
21&  NGC 5813  &	E1-2	  &    0.00658 &  11.45   &  	9517  &  	99	&  	98.1    & BGG	\\
22&  NGC 5846  &        E0-1	  &    0.00572 &  11.05   &  	788   &  	30	&  	23      & BGG   \\
23&  NGC 5866  &        S$0_3$	  &    0.00224 &  10.74   &  	2879  &  	34	&  	29.7    & BGG	\\
\hline                                                                                                            
\end{tabular}}
\label{glo1}
\end{center}
\footnotesize
\begin{flushleft}
{Notes: col.2 - galaxy identification; col.3 - morphological type; col.4 - redshift; col.5 - absolute magnitude in B band; col.6  - \textit{Chandra} observation identification number; Col. 7 - exposure time; Col. 8 - good time interval (GTI) after light curve filtering; \& Col. 9 - environments of the target galaxies; where GG: member of the group, FG: Field galaxy, CG: member of cluster, LG: member of loose group, BGG: brightest galaxy in the group, BCG: brightest cluster galaxy.}
\end{flushleft}
\end{table}

Traditionally, ETGs were regarded as simple structures, devoid of gas and dust. However, the classical notion of being dry, passively evolving systems have amply changed with the employment of both ground and space-based telescopes across the electromagnetic spectrum. Extensive studies conducted on the content of ISM in early-type galaxies using the multi-wavelength data have established that the ETGs hosts complex, multi-phase ISM (e.g. \cite{1987IAUS..127..135B}, \cite{1989PhDT.........3E},\cite{1989ApJ...336..822K}, \cite{1995AJ....110.2027V}, \cite{1994A&AS..105..341G}, \cite{2007A&A...461..103P}, \cite{2010MNRAS.409..727F}). In particular, about 50-80\% of the ETGs are known to host dust in a variety of morphological forms (see e.g. \cite{1994A&AS..105..341G}, \cite{2007A&A...461..103P}, \cite{2010MNRAS.409..727F}, \cite{2012MNRAS.422.1384F}). Further, past studies have also demonstrated that dust in these galaxies is having external origin i.e., accreted by the system either through interaction or merger like event (\cite{1988MNRAS.234..733B}, \cite{1994MNRAS.271..833G}, \cite{2007A&A...461..103P}, \cite{2009arXiv0901.1747P}, \cite{2012NewA...17..524V}). Given the external origin of the dust, investigation of populations of XRBs in dusty ETGs are important to constrain the formation scenario of this class of galaxies. Therefore, a systematic study of properties of discrete sources in a relatively larger sample of dusty early-type galaxies is called for. 

We have an ongoing program of examining the association of multi-phase ISM in a sample of dusty early type galaxies (D-ETGs) selected from different environments i.e, isolated regions, fields, groups and clusters. Analysis of high resolution multi-wavelength data on these galaxies has confirmed the spatial correspondence between the dust and the ionized gas in a large fraction of galaxies and in some cases with the X-ray emitting region too, pointing towards their common origin (\cite{2012NewA...17..524V}, \cite{2012MNRAS.421..808P}). During this study we found that the D-ETGs also host a significant number of discrete sources whose populations and characteristics are believed to vary as a function of environment of the host galaxy. This paper presents a systematic study of the spectral properties of discrete sources detected within the \D (isophote of the 25.0 B-mag arcsec$^{-2}$ brightness level; \cite{1991Sci...254.1667D}) regions of the sample galaxies. In addition to this, an attempt is also made to classify these sources on the basis of their X-ray colors and also to disentangle their contribution to the total X-ray luminosity of the host. This paper is structured as follows: Section 2 describes the sample selection, X-ray observations and the steps involved in the data preparation. Results derived from this analysis are presented in Section 3, whereas Section 4 discusses some of the important results. All the distance dependent estimates are based on the H$_0$ = 70 km s$^{-1}$ Mpc$^{-1}$. 

\section{Observations and Data Preparation}
We have constructed a heterogeneous sample of 23 nearby early-type galaxies on the basis of their \lq\lq{dustyness}\rq\rq. Care was taken that the objects represent different environmental conditions i.e., isolated regions, groups and clusters. Here, we considered the objects with radial velocities $\le$ 5000 km/s that were observed for at least about 20 ks by the \textit{Chandra} telescope. The global properties of the target galaxies along with details of their \textit{Chandra} observations are given in Table~\ref{glo1}. Most of the target galaxies selected for the present study were observed with the ACIS-S detector, except for NGC 1395 and NGC 3607 where ACIS-I was employed. 

The data products on the target galaxies were uniformly processed using the \emph{Chandra} Interactive Analysis of Observations software package (CIAO) v4.2.0 and the latest calibration files provided by the \textit{Chandra} X-ray data Centre (CALDB v4.3.1). Periods of high background count rates were identified by examining the light curves constructed after extracting X-ray photons from the outer regions of the ACIS chip and were filtered out using the 3$\sigma$ clipping algorithm \emph{lc$\_$sigma$\_$clip} available within CIAO. The resultant good time intervals (GTI) for each of the target galaxies are given in column 8 of Table~\ref{glo1}. These light curve filtered event files were then used for the further analysis. 
\begin{figure}
\includegraphics[width=90mm,height=80mm]{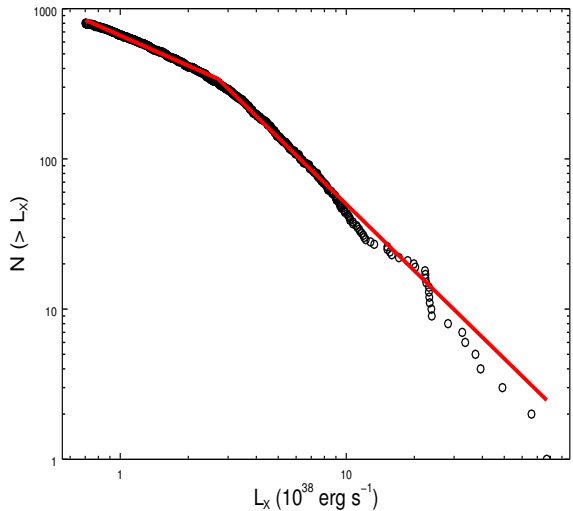}
\centering
\caption{\label{xlf} The combined X-ray luminosity function of all the 996 discrete sources detected within \D region of 23 dusty early-type galaxies. The solid line shows the best-fitted broken power-law showing a break at \s 2.71\tim \te \lum.} 
\end{figure}

Point sources present within the optical \D ellipse of each of the target galaxy were detected using a wavelet source detection algorithm \textit{wavdetect} adopting a detection threshold of 10$^{-6}$. For this we ran \emph{wavdetect} on the full-band (0.3-10.0\,keV) \emph{Chandra} images on the target galaxies with a ``$\sqrt{2}$ sequence'' of wavelet scale increasing from 1 to 32 pixels. In order to minimize contamination from unrelated foreground/background sources, we restricted our source detection to the \D\, ellipse. This analysis enabled us to resolve a total of 996 discrete point-like sources in the target galaxies (see Table~\ref{txrb}). After detecting the sources and finding their positions, we extracted X-ray counts from circular regions centered on individual sources along with their local backgrounds. The background regions were chosen to cover an area at least 3 times the source's extraction region. 

\section{Results}
\subsection{X-ray Luminosity Function}
The X-ray luminosity function (XLF) of the discrete sources is an effective tool for investigating properties of the populations of XRBs (e.g., \cite{2003MNRAS.339..793G}). To study the luminosity distribution of the XRBs in E and S0 galaxies, XLFs have been extensively obtained for a large number of galaxies using \textit{Chandra} data (\cite{2003ApJ...587..356I}, \cite{2003ApJ...585..756J}, \cite{2003ApJ...599..218S}, \cite{2008ApJ...689..983H}, \cite{2004ApJ...611..846K}). The functional shapes and presence of breaks in the XLFs have been examined by fitting theoretical models to them. A broken power law with the break at \s\,(2 - 5) \tim\, \te\, \lum, analogous to the Eddington luminosity of a neutron star accreting binary, are found to well represent the functional shapes of the XLFs for population of XRBs in ETGs  (e.g., \cite{2000ApJ...544L.101S}, \cite{2002ApJ...574..754F}, \cite{2004ApJ...611..846K}, \cite{2009arXiv0903.3123V}). 

With a view to study the luminosity distribution of the population of discrete sources, we derived cumulative X-ray luminosity function (XLF) of all the 996 discrete sources detected within \D\, of the sample galaxies. Before this, we fitted the cumulative spectra of all the sources within a given galaxy and estimated the appropriate energy conversion factors in units of erg/counts. Then the count rates derived in the energy range 0.3-10.0 keV for individual sources were converted into luminosities  assuming that all the sources within a galaxy are at the same distance as that of their host. The resultant unabsorbed X-ray luminosities of individual sources are found to lie in the range between 1.1 \tim\, \ts\, \lum to 7.7 \tim\, \tn\, \lum. 
 
\begin{figure*}
\includegraphics[width=75mm,height=65mm]{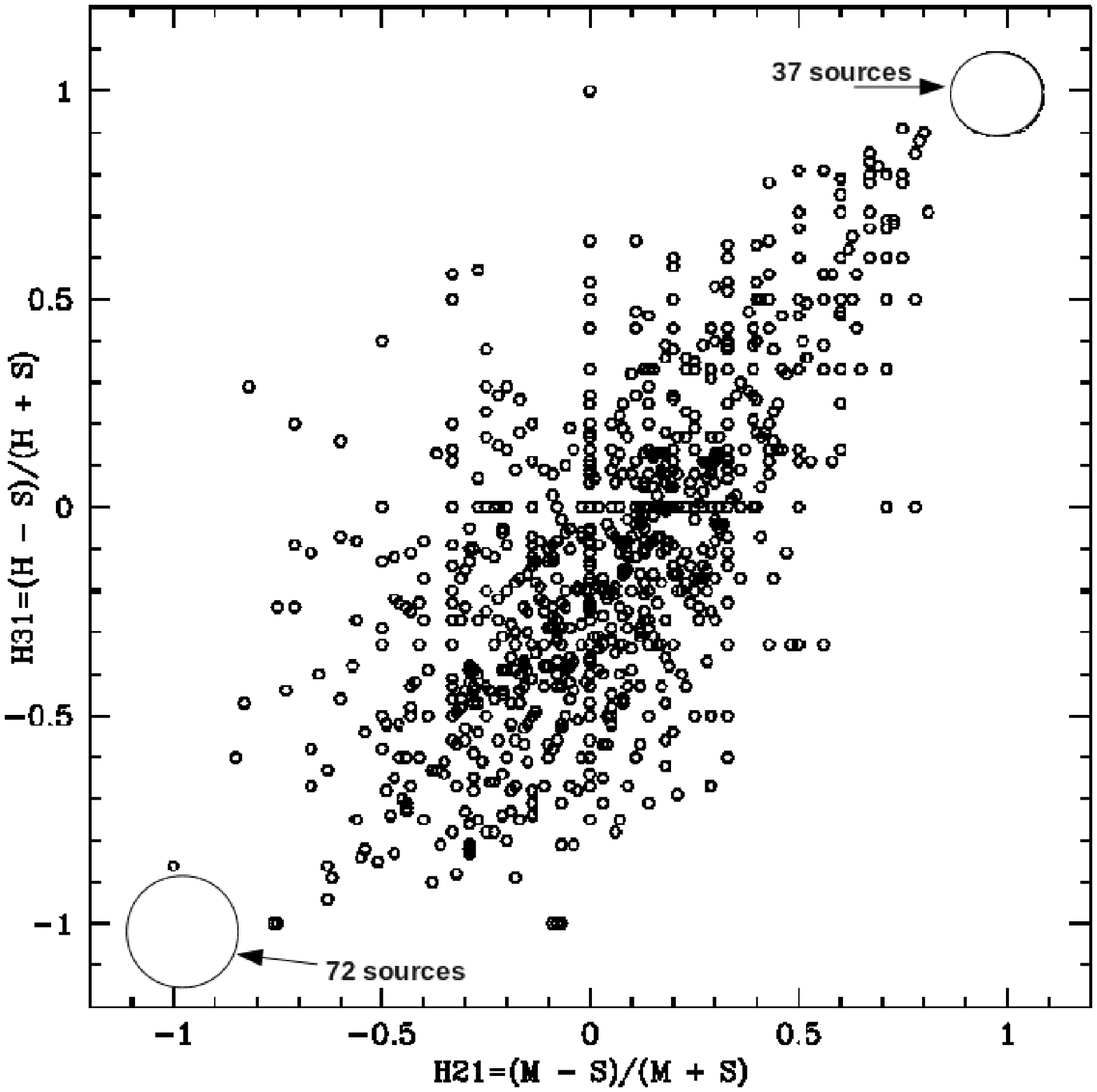}
\includegraphics[trim=0.01cm 0.1cm 0.01cm 0cm, clip=true, width=75mm,height=65mm]{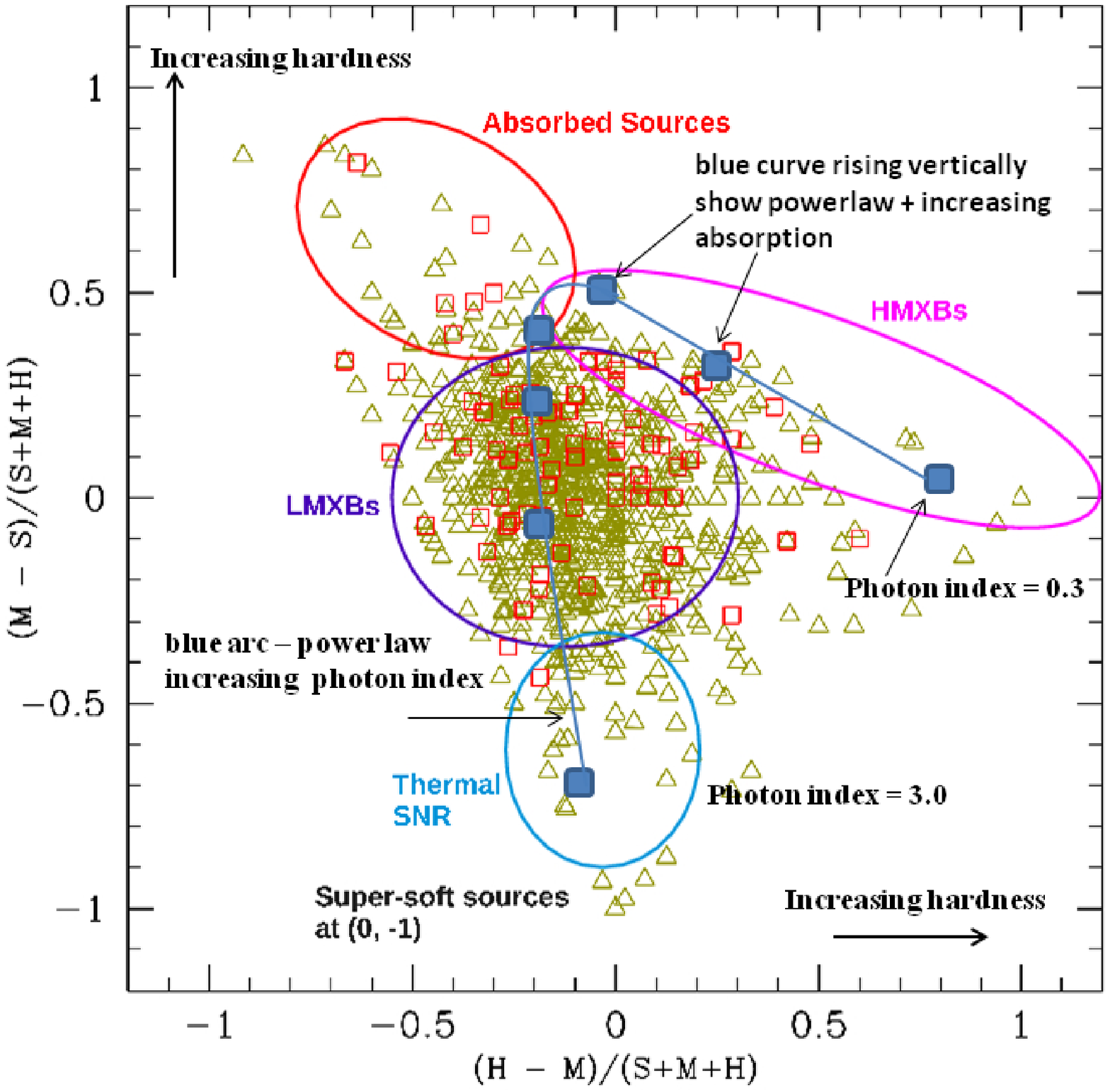}
\centering
\caption{\label{hr1} \textit{left panel}: X-ray hardness ratio plot (H21 vs. H31) of the 996 discrete sources detected within $D_{25}$ ellipses of 23 dusty ETGs from the present sample. Here, H21=(M-S)/(M+S) and H31=(H-S)/(H+S), where S, M and H are the X-ray counts in soft (0.3-1.0\,keV), medium (1.0-2.0\,keV) and hard (2.0-10.0\,keV) bands, respectively. The 72 sources near (-1, -1) are super-soft or very soft sources, while the 37 sources near (+1, +1) may be unrelated background objects. \textit{right panel}: The modified X-ray color-color plot of all the 996 sources. This plot is between hard (HC) versus soft color (SC) of the discrete sources (following \cite{2003ApJ...595..719P}). X-ray hard color is defined as $HC=(H-M)/T$ and X-ray soft color as $SC=(M-S)/T$ where T=(S+M+H). The golden triangles are sources from the dusty elliptical galaxies, while red squares are those from the dusty lenticular galaxies. The blue curve delineates the power law component of increasing photon index.}
\end{figure*}

Figure~\ref{xlf} presents the combined luminosity distribution of all the sources detected within \D of the target galaxies and shows a "knee" in the overall shape. Therefore, we fitted a broken power law to the combined LMXB XLF with a break at (2.71 \pms 0.03) \tim\te\,\lum, near the Eddington limit of an accreting neutron star and may be related to the transition in the XLF between neutron star and black hole binaries (\cite{2006ARA&A..44..323F}). The best fit parameters determined from the maximum-likelihood method, resulted in the negative differential logarithmic slope values of 0.67\pms0.08 and 1.47\pms0.02 before and after the break, respectively. A dearth of very luminous sources in the sample galaxies is evident in this figure. Our results are consistent with those reported by \cite{2003ApJ...587..356I}  for fitting the cumulative XLF to LMXBs in 15 elliptical galaxies and by \cite{2006ApJ...653..207D} in 18 early-type galaxies. Similar results were also reported by \cite{2004MNRAS.349..146G} for four-early type galaxies and by several other authors (\cite{2000ApJ...544L.101S}, \cite{2001ApJ...556..533S}, \cite{2001ApJ...552..106B}, \cite{2002ApJ...574L...5K}, \cite{2004ApJ...613L.141G}, \cite{2004ApJ...611..846K}, \cite{2004MNRAS.349..146G}, \cite{2008ApJ...689..983H}) for point sources detected in individual galaxies. 
\subsection{Hardness ratio}
In addition to the conventional method of plotting XLFs, X-ray color plot for the resolved sources was found to be a potential way of investigating spectral properties of individual sources. The unique spectral color of each class of the XRBs determines its position in the X-ray color plot and hence enable us to classify the sources. This technique has an advantage that it can be employed even if the counts from individual sources are not sufficient to perform the spectral fit and have been used extensively for delineating properties of individual sources (\cite{2001A&A...373..438H}, \cite{2003ApJ...595..719P}, \cite{2001ApJ...556..533S}, \cite{2002ApJ...570..152I}, \cite{2006ARA&A..44..323F}). 

\begin{figure*}
\includegraphics[width=75mm,height=65mm]{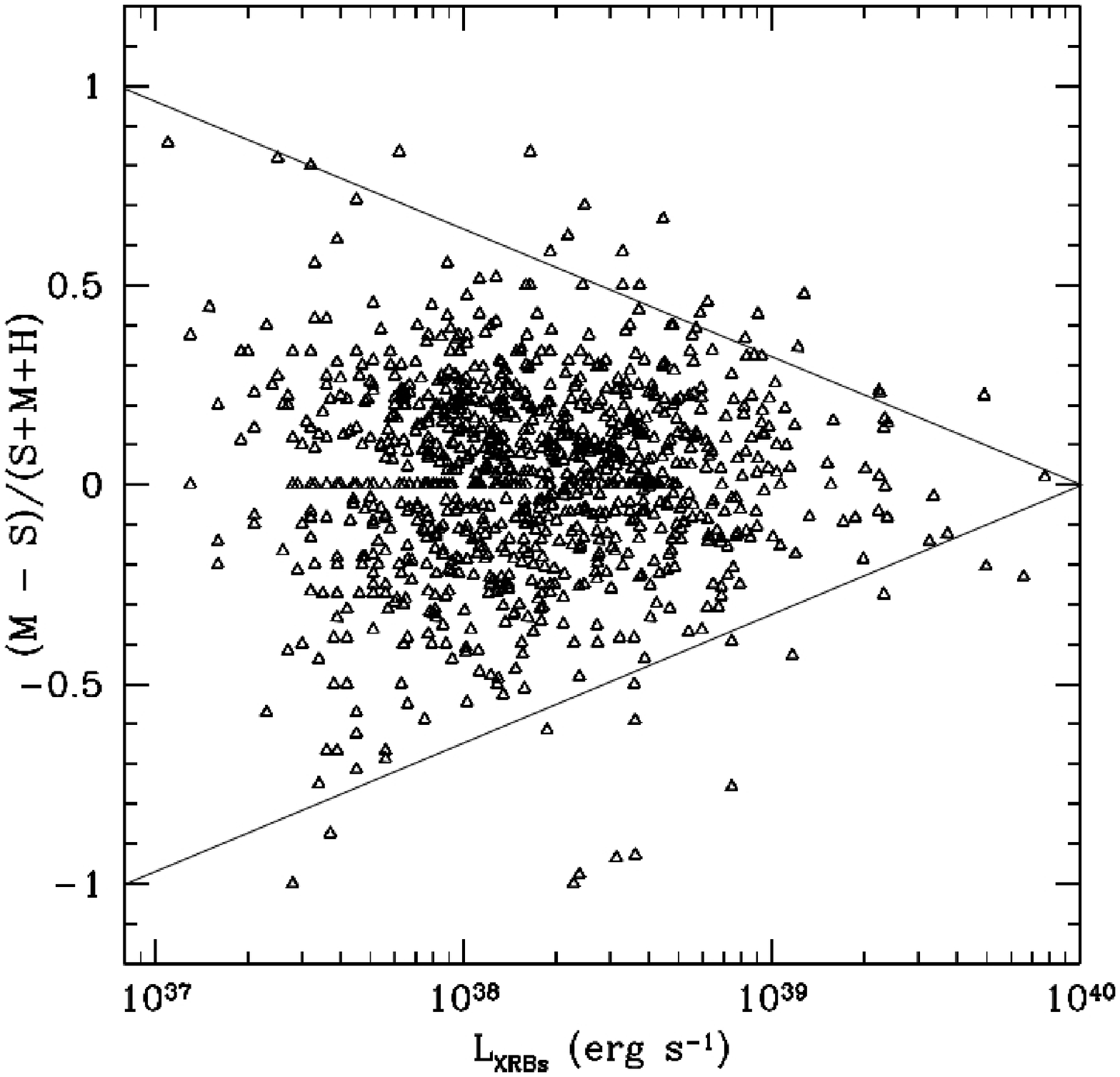}
\includegraphics[width=75mm,height=65mm]{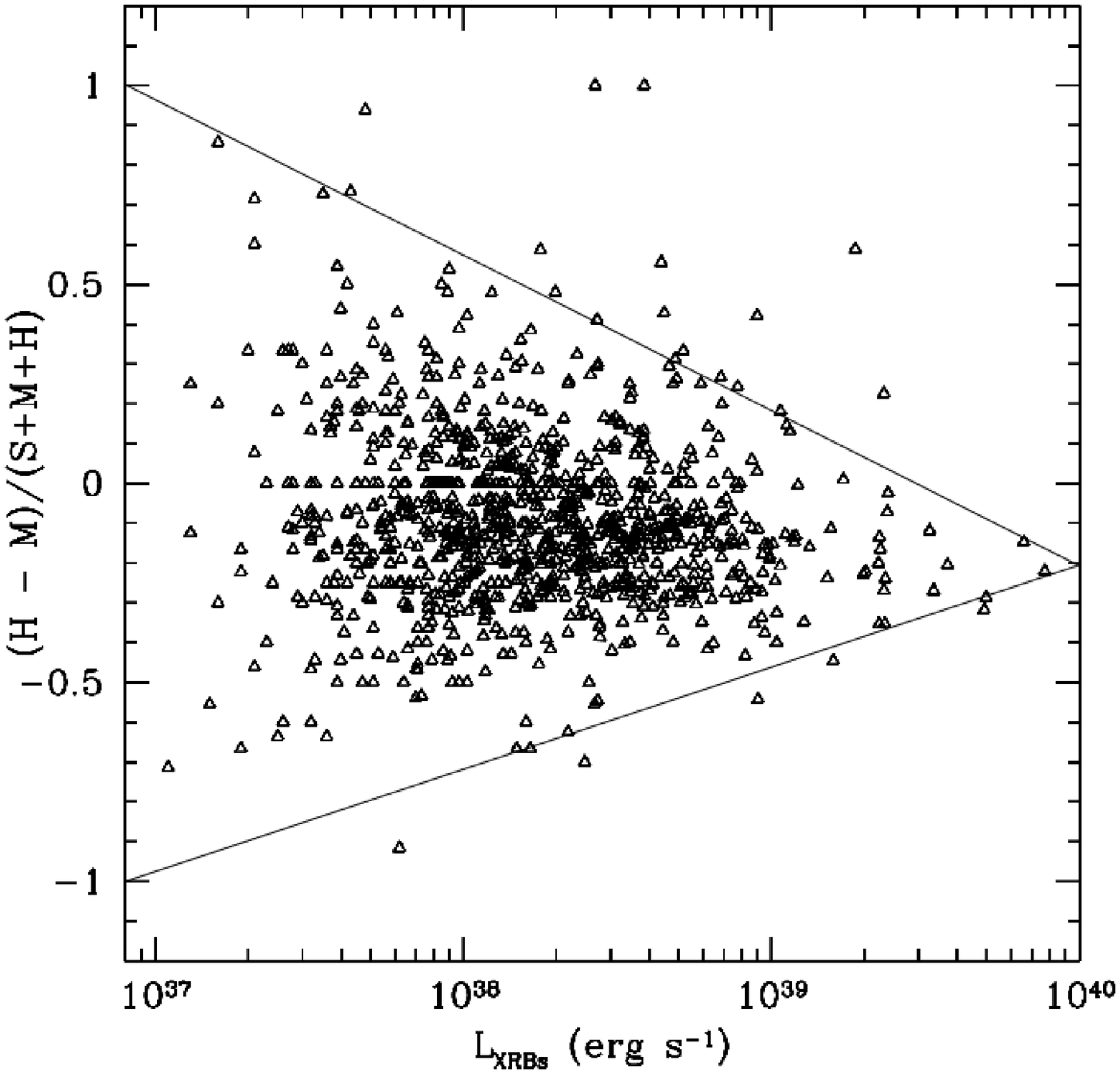}
\centering
\caption{\label{hrlx} The X-ray soft (left panel) \& hard (right panel) colors of individual sources plotted against their luminosities. Both these figures reveal a dearth of the XRBs with \lx $>$ 5\tim\tn\, \lum. }
\end{figure*}

Following the procedure outlined by \cite{2000ApJ...544L.101S}, we quantified hardness ratios of individual sources by extracting source counts in three different energy bands: soft (S, 0.3 - 1.0 keV), medium (M, 1.0 - 2.0 keV), and hard (H, 2.0 - 10.0 keV). We then plot H21=(M-S)/(M+S) versus H31=(H-S)/(H+S) for all the 996 sources. The resultant color-color diagram is shown in Figure~\ref{hr1} (left panel).  From this figure it is apparent that all the sources lie along the diagonal band from (H21, H31) = (-0.95, -0.90) to (+1.00, +1.00) with a cluster at about (-0.06, -0.07), consistent with the observations by (\cite{2002ApJ...570..152I}, \cite{2003ApJ...585..756J}, \cite{2004ApJ...600..729R}, \cite{2000ApJ...544L.101S}, \cite{2001ApJ...556..533S}, \cite{2001ApJ...552..106B}). Among the 996 sources, 72 have their hardness ratios close to (-0.95, -0.90), implying that these sources essentially have no counts above 1\,$keV$ and most likely are the super-soft or very-soft sources (\cite{2001ApJ...556..533S}). A blackbody spectrum with a temperature of 0.1\,$keV$ and Galactic absorption about 2.0\tim\, 10$^{20}$ \cms\, would give hardness ratio close to \s\,(-0.97, -1.00) (\cite{2001ApJ...556..533S}). There are about 37 sources whose hardness ratios are close to (+1, +1) and are hardest among all the sources. A power-law spectrum with a photon index $\Gamma$\s\,1.5 and an absorbing column density of N$_H$ \s\, 10$^{22}$ \cms\,  would give hardness ratio close to \s\,(+0.88, +0.92) (\cite{2001ApJ...556..533S}).  

Though the conventional X-ray color plot is a useful tool for studying properties of LMXBs, it fails to classify the sources in a statistical sense for studying their populations. To gain insight in to the details of the discrete sources, we have derived modified colors following the procedure outlined by \cite{2003ApJ...595..719P}. Here, the modified colors are defined as HC = (H-M)/T (hard color) and SC = (M-S)/T (soft color), where S, M, H and T are X-ray counts in soft (0.3-1.0\,keV), medium (1.0-2.0\,keV), hard (2.0-10.0\,keV) and total (0.3-10.0\,keV) bands, respectively. Figure~\ref{hr1} (right panel)  shows the modified X-ray color-color diagram for all the 996 discrete sources and exhibit significantly different color values for sources of different populations, providing a relatively assumption-free tool for the separation of the sources (\cite{2006ARA&A..44..323F}). From this figure it is apparent that, about 63$\%$ of the sources in the dusty early-type galaxies are dominated by the low mass X-ray binaries with their color values lying between -0.4 and +0.4 and are identical to those seen in the bulge of the Galaxy. 

This figure shows a population of the soft and very soft sources with soft X-ray color between -0.3 to -0.9 and are similar to those seen in the disks of star forming galaxies (\cite{2003ApJ...595..719P}). The very-soft sources occupy positions below the thermal SNRs and are found to be associated with the star forming regions. As the rate of star formation in these galaxies is very low, therefore one expects very few numbers of sources of this kind. In the present study we have detected about 20 such soft and very-soft sources in 23 D-ETGs. There are few sources with SC=-1.0, essentially having no counts above 1\,keV, and represent classical super-soft sources. This figure also shows few sources with HC $>$ 0.0 with larger flux in harder band and are HMXBs. The HMXBs are short lived objects and appear in the active star forming regions of late-type galaxies.  A handful of sources with SC$>$ +0.40 are also evident in this figure and probably represent the heavily absorbed harder sources unrelated to the target galaxies. 

Figure~\ref{hrlx} represents the plots between the X-ray soft (left panel) and hard colors (right panel) of the sources against their X-ray luminosities. Both these plots confirm the dearth of high luminosity XRBs with L$_X > $ 5\tim \te\,\lum  in the dusty early-type galaxies. The larger scatter evident in the X-ray colors of the sources with L$_X < $ 2\tim\, 10$^{38}$ \lum\, is due to the nature of the sources.

\begin{figure}
\includegraphics[trim=0.01cm 0.1cm 0.01cm 0cm, clip=true, width=75mm,height=65mm]{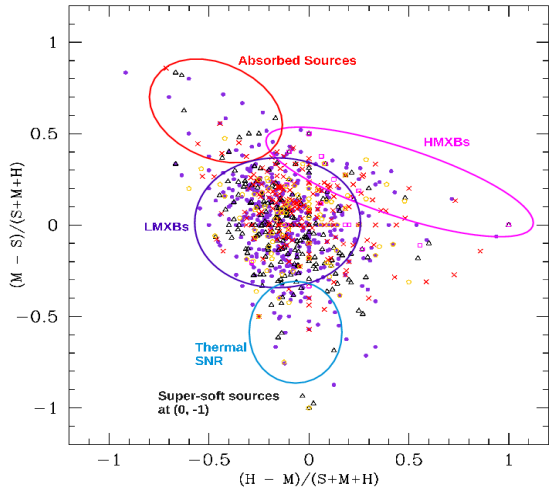}
\centering
\caption{\label{hren} Environment dependent X-ray color-color plot for all the discrete sources resolved within the target galaxies. Different color codes are: BGG: black triangle; GG: red cross; GC: blue-violet filled circle; BCG: magenta open circle; FG: gold pentagon. }
\end{figure}

\begin{table*}
\caption{Spectral properties of the discrete sources}
\scriptsize{
\hspace{-1cm}\begin{tabular}{@{}|cccccccc|r@{}}
\hline
Obj.  &	No. of srcs. &  N$_H$ &  $\Gamma$ & L$_{XRBs}$  & L$_{tot}$ & XRBs contri. &	$\chi^2$/dof \\
      &     &  (10$^{20}$\,\cms) &  &(10$^{40}$\lum) & (10$^{40}$\lum)  & ($\%$) &              \\
  (1)    &  (2)   &  (3) & (4) &(5) & (6) & (7) & (8)                \\
\hline
NGC 1395 &	24   &	1.66  &	1.84\pms0.14  &	1.46\pms0.14 & 5.45  & 26.79   &   22.06/23 = 0.96  \\
NGC 1399 &	22   &	1.50  &	1.97\pms0.12  &	1.07\pms0.17 & 10.08 & 10.62   &   7.35/9 = 0.82      \\
NGC 1404 &	18   &	1.50  &	1.10\pms0.12  &	1.24\pms0.14 & 21.18 & 5.85    &   25/20 = 1.25       \\
NGC 1407 &	66   &	5.40  &	1.51\pms0.06  &	2.57\pms0.17 & 12.68 & 20.27   &   97.15/76 = 1.27	 \\
NGC 2768 &	45   &	4.13  &	1.46\pms0.06  &	0.97\pms0.05 & 1.77  & 54.8    &   59.40/76 = 0.78    \\
NGC 3377 &	22   &	2.86  &	1.49\pms0.11  &	0.14\pms0.01 & ---   &  ---    &   39.27/38 1.03   \\
NGC 3379 &	43   &	2.91  &	1.74\pms0.09  &	0.39\pms0.02 & 0.48  & 81.25   &   36.37/33=1.10   \\
NGC 3585 &	31   &	5.32  &	1.57\pms0.09  &	0.71\pms0.04 & 1.02  & 69.61   &   39.85/46 = 0.86 \\
NGC 3607 &	10   &	1.36  &	1.46\pms0.27  &	0.16\pms0.02 & ---   & ---     &   30.02/24= 1.25   \\
NGC 3801 &	06   &	1.97  &	1.69\pms0.37  &	0.24\pms0.04 & 2.69  & 8.92    &   8.05/9 = 0.90       \\
NGC 3923 &	18   &	6.26  &	1.70\pms0.09  &	2.51\pms0.14 & 8.66  & 28.98   &   43.09/32 = 1.32	\\
NGC 4125 &	33   &	1.74  &	1.78\pms0.06  &	1.04\pms0.05 & 2.72  & 38.24   &   73.41/85 = 0.86  \\
NGC 4278 &	104  &	2.07  &	1.44\pms0.04  &	0.31\pms0.01 & 0.81  & 38.27   &   70.72/80=0.88   \\
NGC 4365 &	97   &	1.70  &	1.58\pms0.06  &	1.28\pms0.07 & ---   & ---     &   87.45/81 = 1.08 \\
NGC 4374 &	51   &	2.99  &	1.87\pms0.05  &	1.58\pms0.06 & 8.49  & 18.61   &   125.76/103 = 1.22 \\
NGC 4473 &	22   &	2.39  &	1.67\pms0.17  &	1.06\pms0.09 & 1.92  & 55.21   &   17/18 = 0.95    \\
NGC 4494 &	22   &	1.42  &	1.78\pms0.13  &	0.69\pms0.07 & ---   & ---     &   25.55/27 = 0.95 \\
NGC 4552 &	85   &	2.62  &	1.36\pms0.04  & 0.16\pms0.05 & 0.44  & 36.36   &   115.15/89 = 1.29 \\	
NGC 4649 &	142  &	2.10  &	1.70\pms0.04  &	2.52\pms0.05 & 11.06 & 22.78   &   168.40/99 = 1.70  \\	
NGC 4697 &	68   &	2.03  &	1.66\pms0.05  &	1.54\pms0.07 & 2.16  & 71.3    &   128.05/107 = 1.19 \\
NGC 5813 &	27   &	4.37  &	1.84\pms0.05  &	2.02\pms0.10 & ---   &  ---    &   121.70/109 = 1.11 \\
NGC 5846 &	14   &	3.97  &	1.75\pms0.22  &	0.50\pms0.06 & 23.5  & 2.13    &   13.16/11 = 0.93 \\
NGC 5866 &	26   &	1.38  &	1.83\pms0.14  &	0.10\pms0.07 & 0.27  & 37.04   &   24.53/24 = 1.02 \\
\hline
\end{tabular}}
\footnotesize
\begin{flushleft}
{Note: Col 1. galaxy identification; col 2. No. of sources resolved within \D\, region of the galaxy; col 3. Galactic hydrogen column density; col 4. power-law photon index $\Gamma$; col 5. cumulative X-ray luminosity of all the sources resolved within \D\, of the target galaxy; col 6. total (diffuse+resolved) X-ray luminosity; col 7. - contribution of X-ray binary sources within \D; col 8. $\chi^2$ per degree of freedom and goodness of fit}
\end{flushleft}
\label{txrb}
\end{table*}
\subsection{Spectral properties of LMXBs in different environments}
As count rates were not enough to perform spectral analysis of individual sources, therefore, we analyzed the combined spectrum of all the sources within a galaxy. For this, a cumulative 0.3-10.0 keV spectrum of all the resolved sources was extracted using the CIAO task \textit{acisspec}. We first tried with a two component (MEKAL + bremsstrahlung) model, with their normalizations as free parameters. This gave a reasonably good fit for some of the objects; however, due to very weak flux in the soft band it gave an upper limit in majority of the cases. We then removed the soft component (MEKAL) in the fit. This resulted in the fit just as good after including it, implying that the resolved sources have no soft component in their spectra.  Next, we allowed the absorbing column to vary with a view to see if there was evidence of any excess absorption beyond the Galactic column. However, allowing the absorbing column to vary does not improve the fit. Therefore, we fixed the hydrogen column at the Galactic value and replaced the harder component, thermal bremsstrahlung, by a power-law with photon spectral index of $\Gamma$. After including the power-law quality of fit was improved, therefore, we adopted this as the best fit model. The best-fitted power-law indices for resolved sources in individual galaxies along with their combined luminosities and goodness of fit are listed in Table~\ref{txrb} and are consistent with those fitted for the emission from LMXBs in 15 early-type galaxies by \cite{2003ApJ...587..356I} and for 18 early-type galaxies by \cite{2006ApJ...653..207D}. 

\section{Discussion and Conclusions}
We have presented the results based on a systematic analysis of the \textit{Chandra} X-ray observations of a heterogeneous sample of 23 dusty early-type galaxies selected from different environments. This study has enabled us to resolve a total of 996 discrete sources within optical $D_{25}$ region of target galaxies. Some of the galaxies from this sample, e.g. NGC 4278, NGC 4365, NGC 4649 etc. are found to host a significantly large number of sources, while others host very few. Further exploration revealed that the systems hosting a larger population of XRBs are either field galaxies or belong to the loose groups, whereas those belonging to rich groups or clusters host very few sources. 

\begin{figure}
\includegraphics[width=70mm,height=60mm]{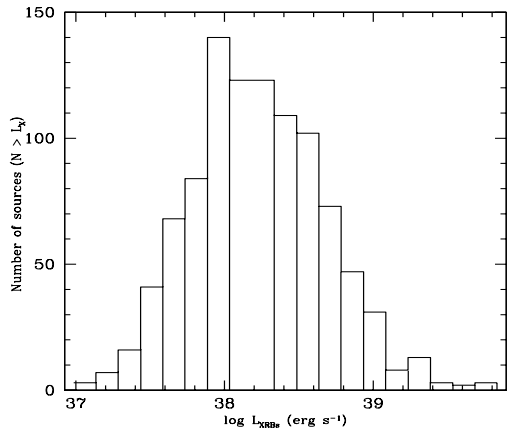}
\centering
\caption{Histogram showing the luminosity distribution of all the 996 resolved sources from the present study.}
\label{histo}
\end{figure}

Intrinsic luminosities of the individual sources are found to lie in the range between 1.14 \tim\,\ts\,\lum\, to 7.7 \tim\,\tn\,\lum, with the higher limit reached by the Ultra Luminous X-ray Sources (ULXs) which are very few in number. Out of 996, about 63\%\,have their properties similar to the normal LMXBs. About 15-20\%\,of the sources with luminosities of a few \tim \ts\, \lum\, are either super-soft or very-soft sources or heavily absorbed SNRs and are likely to be associated with the star forming regions in the galaxies. Remaining are either ULXs, HMXBs or heavily absorbed harder sources. The histogram showing the luminosity distribution of the resolved sources is given in Figure~\ref{histo} and implies that majority of the sources have emission characteristics similar to the 1.4\Msun\,neutron star accreting LMXBs. As regard to the contribution of the discrete sources to the total X-ray luminosity of the host galaxies, it is found that in some of the galaxies from isolated regions (e.g., NGC 3379) the contribution of the XRBs is very high \s 81\%, while for those from rich clusters (e.g., NGC 5846) it is only 2\%. With a view to examine the environment dependent nature of the XRBs we have plotted the X-ray color-color graph of discrete sources from different galaxies in Figure~\ref{hren}. This figure does not show any obvious structural differences in the nature of the sources representing different environmental conditions. 

The issue of the origin of LMXBs has been debated ever since their discovery in the Milky-Way (\cite{1974ApJS...27...37G}). A number of galaxies has been studied to this date with \textit{Chandra} to understand the origin of the LMXBs. Combined optical and X-ray studies of a large sample of galaxies have revealed that 20\% - 70\% of the LMXBs in ETGs are associated with the globular clusters (GCs; \cite{2001ApJ...556..533S}, \cite{2003ApJ...589L..81K}, \cite{2003ApJ...599..218S}, \cite{2004ApJ...612..848H}, \cite{2004ApJ...600..729R}). However, a substantial number of LMXBs have also been detected in the fields of the early-type galaxies. If all the LMXBs were formed in the GCs then regardless of where the LMXBs are distributed, the combined X-ray luminosity of all the LMXBs in a galaxy should scale with the number of globular clusters hosted by it  (\cite{2006ESASP.604..455I}). Therefore, to check their relevance with the GCs we explored a similar correlation between the combined X-ray luminosity of the LMXBs detected within a galaxy versus its GC specific frequency (S$_N$; taken from \cite{2008ApJ...681..197P},\cite{1997AJ....113.1652F},\cite{2004ApJ...611..846K}). The resultant plot showed a very weak weak correlation between the two quantities, implying that origin of at least a fraction of LMXBs in the dusty-ETGs could be through the evolution of primordial binaries. 

\begin{figure}
\includegraphics[width=70mm,height=65mm]{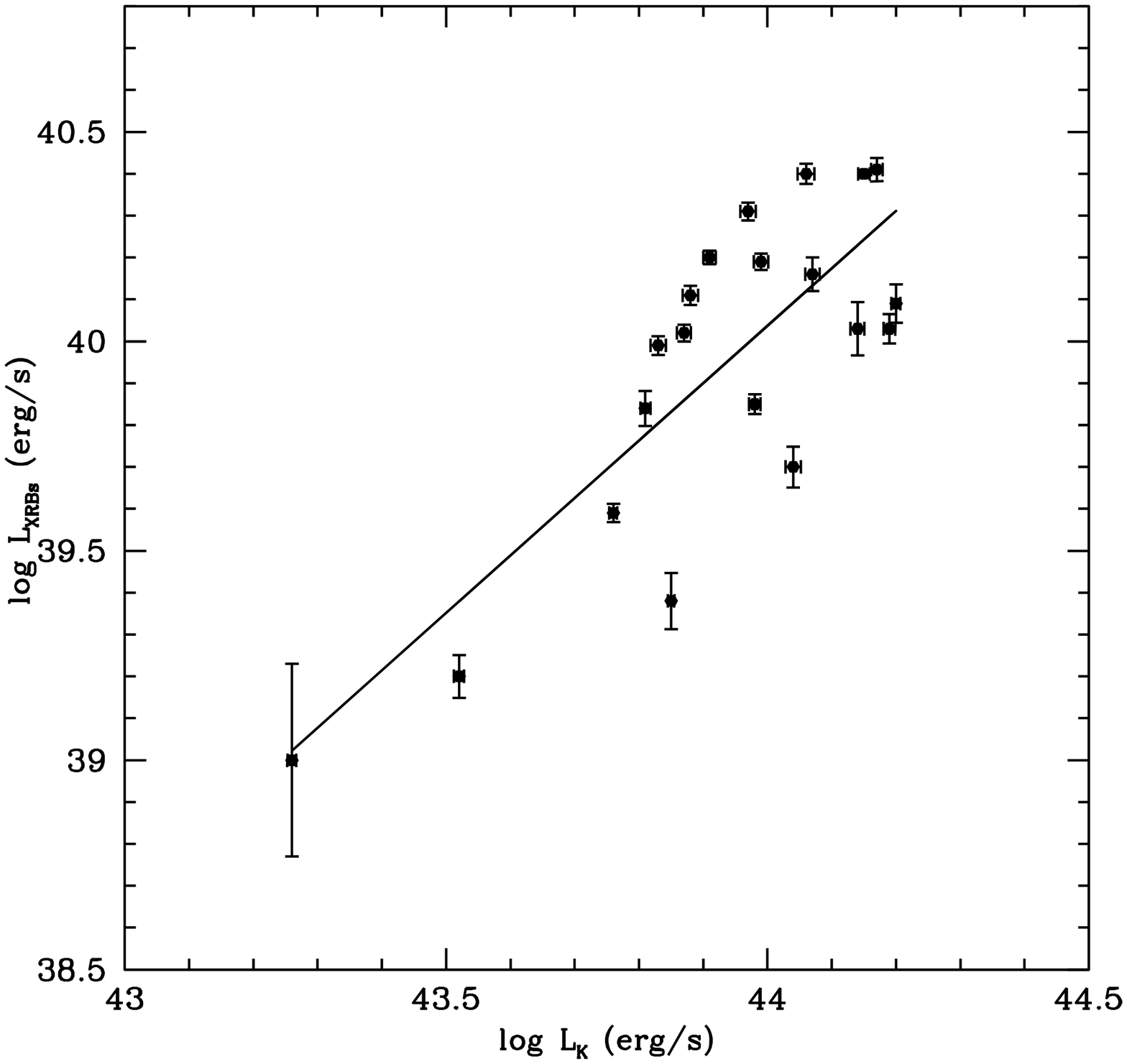}
\includegraphics[trim=0.5cm 0cm 0.5cm 0cm, clip=true, width=70mm,height=65mm]{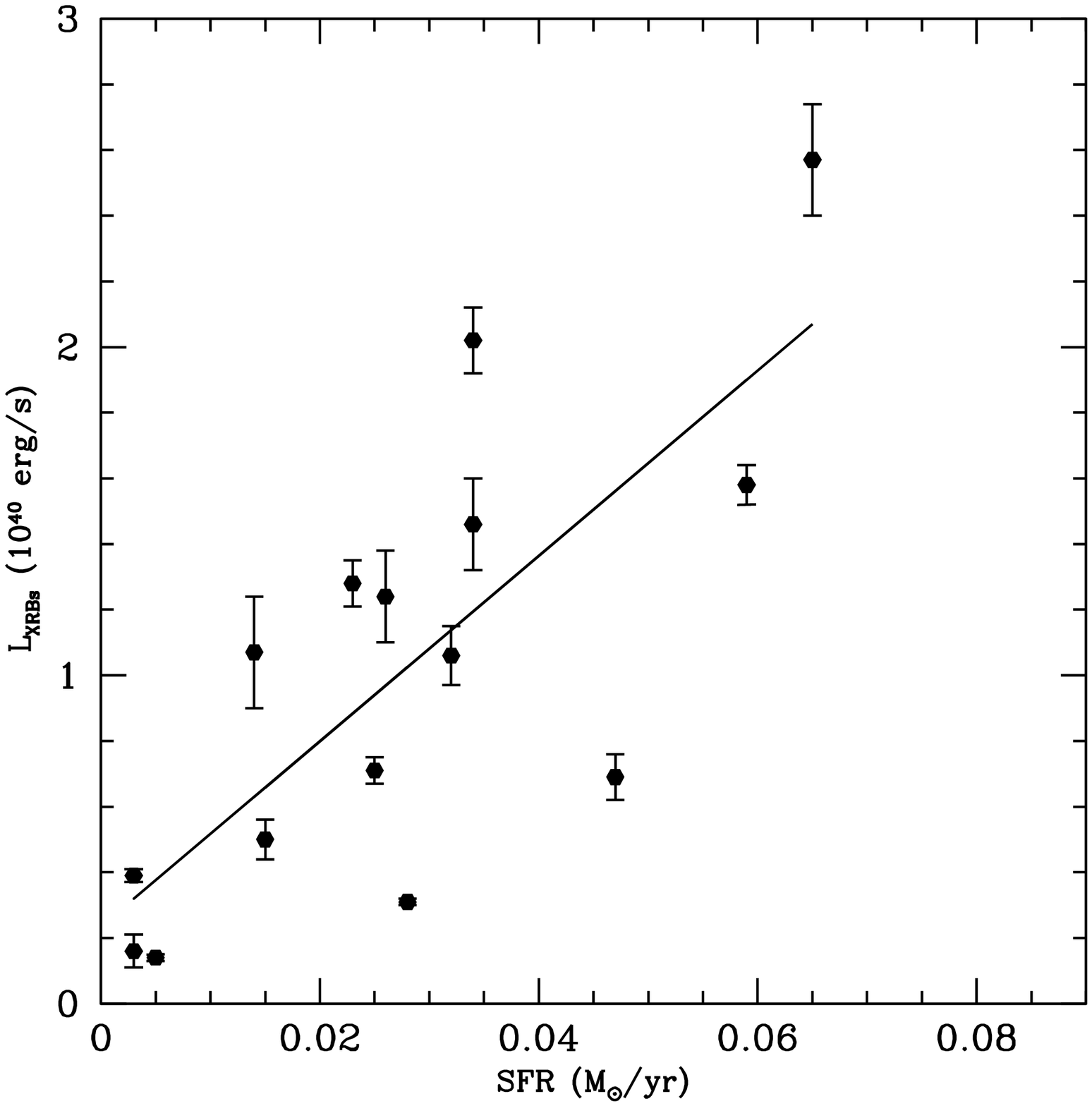}
\centering
\vspace{-0.7cm}
\caption{\label{lxk} Correlation between the combined X-ray luminosity of the resolved sources within a galaxy versus their K$_s$ band luminosities (left panel) and star formation rate (SFR, right panel). }
\end{figure}

After realizing that a significant number of LMXBs are field sources, we explored their correlation with the stellar content of the host galaxies.  The D-ETGs being dominated by the old, long-lived stellar populations, near-IR luminosities provide a better proxy for the stellar mass content of these galaxies. Hence we plotted a correlation between the combined X-ray luminosity of the LMXBs in a galaxy with its K$_s$ band luminosity (from 2MASS data) which is shown in Figure~\ref{lxk} (left panel), giving the best fit slope value equal to 1.34\pms0.33. The large scatter evident in this graph is due to the nature of the X-ray sources and is consistent with those studied previously  (\cite{2006ApJ...653..207D}). As there is growing evidence for ongoing star formation in the D-ETGs, we have also checked the correlation between L$_{XRB}$ and star formation rate (SFR; estimated from the IR luminosities, \cite{1998ARA&A..36..189K}) which is given in Fig.~\ref{lxk} (right panel). 

As merging has been suggested as a key factor for the formation of the dusty early-type galaxies (\cite{1987IAUS..127..135B}), one expects a relationship between the age of the D-ETGs and the population of the LMXBs. \cite{2010ApJ...721.1523K} have made an attempt to examine the structural differences in the nature of populations of LMXBs by comparing the XLFs for a sample of young, post-merger elliptical galaxies with that of the old elliptical galaxies. This analysis has enabled them to find a significant difference in the two XLFs particularly at higher X-ray luminosities. The XLF for the young elliptical galaxies exhibits a considerably flatter slope at higher X-ray luminosities (\lx$\ge$5\tim\te \lum) compared to that of the old sample. This means, elliptical galaxies formed recently through the merger like event host relatively larger number of luminous LMXBs (\cite{2010ApJ...721.1523K}).  The cumulative XLF plotted in the present study shows a dearth of high luminosity XRBs (Fig.~\ref{xlf}), implying that probably the sample studied here represents relatively older candidates. Moreover, 10 galaxies from our sample are overlapping with that of \cite{2010ApJ...721.1523K} in which we found a larger number of less luminous LMXBs. 

\section{Acknowledgments}
Authors gratefully acknowledge constructive comments on the manuscript by the anonymous referee, that enabled us to improve it significantly. NDV is thankful of Dr Sean Farrell for his suggestions/inputs on the manuscript. This work is supported by UGC, New-Delhi under the major research project F.No. 36-240/2008 (SR-). We gratefully acknowledge the use of computing and library facilities of the Inter-University Center for Astronomy and Astrophysics (IUCAA), Pune, India. MBP gratefully acknowledge support by INSPIRE Fellowship DST, New Delhi F. No: IF10179. This work has made use of data from the Chandra data archive, NASA’s Astrophysics Data System(ADS), NASA/IPAC Extragalactic Database (NED), provided by CXC.


\def\aj{AJ}%
\def\actaa{Acta Astron.}%
\def\araa{ARA\&A}%
\def\apj{ApJ}%
\def\apjl{ApJ}%
\def\apjs{ApJS}%
\def\ao{Appl.~Opt.}%
\def\apss{Ap\&SS}%
\def\aap{A\&A}%
\def\aapr{A\&A~Rev.}%
\def\aaps{A\&AS}%
\def\azh{AZh}%
\def\baas{BAAS}%
\def\bac{Bull. astr. Inst. Czechosl.}%
\def\caa{Chinese Astron. Astrophys.}%
\def\cjaa{Chinese J. Astron. Astrophys.}%
\def\icarus{Icarus}%
\def\jcap{J. Cosmology Astropart. Phys.}%
\def\jrasc{JRASC}%
\def\mnras{MNRAS}%
\def\memras{MmRAS}%
\def\na{New A}%
\def\nar{New A Rev.}%
\def\pasa{PASA}%
\def\pra{Phys.~Rev.~A}%
\def\prb{Phys.~Rev.~B}%
\def\prc{Phys.~Rev.~C}%
\def\prd{Phys.~Rev.~D}%
\def\pre{Phys.~Rev.~E}%
\def\prl{Phys.~Rev.~Lett.}%
\def\pasp{PASP}%
\def\pasj{PASJ}%
\def\qjras{QJRAS}%
\def\rmxaa{Rev. Mexicana Astron. Astrofis.}%
\def\skytel{S\&T}%
\def\solphys{Sol.~Phys.}%
\def\sovast{Soviet~Ast.}%
\def\ssr{Space~Sci.~Rev.}%
\def\zap{ZAp}%
\def\nat{Nature}%
\def\iaucirc{IAU~Circ.}%
\def\aplett{Astrophys.~Lett.}%
\def\apspr{Astrophys.~Space~Phys.~Res.}%
\def\bain{Bull.~Astron.~Inst.~Netherlands}%
\def\fcp{Fund.~Cosmic~Phys.}%
\def\gca{Geochim.~Cosmochim.~Acta}%
\def\grl{Geophys.~Res.~Lett.}%
\def\jcp{J.~Chem.~Phys.}%
\def\jgr{J.~Geophys.~Res.}%
\def\jqsrt{J.~Quant.~Spec.~Radiat.~Transf.}%
\def\memsai{Mem.~Soc.~Astron.~Italiana}%
\def\nphysa{Nucl.~Phys.~A}%
\def\physrep{Phys.~Rep.}%
\def\physscr{Phys.~Scr}%
\def\planss{Planet.~Space~Sci.}%
\def\procspie{Proc.~SPIE}%
\let\astap=\aap
\let\apjlett=\apjl
\let\apjsupp=\apjs
\let\applopt=\ao
\bibliographystyle{mn.bst}
\bibliography{mybib}
\end{document}